# MULTICOLOR OBSERVATIONS OF ASAS 002511+1217.2

GOLOVIN, ALEX[1]; PRICE, AARON[2]; TEMPLETON, MATTHEW[2]; COOK, LEWIS[3]; CRAWFORD, TIMOTHY[4]; HENDEN, ARNE[2]; JAMES, ROBERT[5]; KOPPELMAN, MICHAEL[6]; NELSON, PETER ROBERT[7]; OKSANEN, ARTO[8]; PAVLENKO, ELENA[9]; PICKARD, ROGER[10]; QUINN, NICK[11]; STARKEY, DON RAY[12];

[1] Visiting astronomer of Crimean Astrophysical Observatory, Komunarov 63, #39, Berdyansk, Zaporozhskaja, 71118, UKRAINE; e-mail: astronom_2003@mail.ru

[2] AAVSO, Clinton B. Ford Astronomical Data and Research Center. 25 Birch St., Cambridge, MA., USA; e-mail: aavso@aavso.org

[3] CBA Concord, 1730 Helix Ct., Concord, CA 94518 USA; e-mail:lcoo@yahoo.com

[4] AAVSO, Arch Cape Observatory, ARCH CAPE, OR, USA; e-mail:tcarchcape@yahoo.com

[5] AAVSO, Las cruces, NM, USA; e-mail: rajames@zianet.com

[6] University of Minnesota, MN, USA; e-mail: michael@aps.umn.edu

[7] Victoria 3820, AUSTRALIA; e-mail: pnelson@dcsi.net.au

[8] Nyrola Observatory - Jyvaskylan Sirius ry, Kyllikinkatu 1, FI-40100 Jyvaskyla Muurame, FINLAND; e-mail: arto.oksanen@jklsirius.fi

[9] Crimean Astrophysical Observatory, Nauchny, UKRAINE; e-mail:pavlenko@crao.crimea.ua

[10] British Astronomical Association Variable Star Section, 3 The Birches, Shobdon, Leominster, Herefordshire HR6 9NG, ENGLAND; e-mail: rdp@star.ukc.ac.uk

[11] British Astronomical Association Variable Star Section, Steyning, WEST SUSSEX BN44 3LR, ENGLAND; e-mail: nick@njq.me.uk

[12] DeKalb Observatory - MPC H63, Auburn, Indiana, USA; e-mail:starkey73@mchsi.com

| Name of the object: |
|---|
| ASAS 002511+1217.2 (RX J0025.1+1217, 1RXS J002510.8+121725, 2MASS J00251111+1217121) |

| Equatorial coordinates: | Equinox: |
|---|---|
| R.A.= $00^h 25^m 11^s 087$    DEC.= $+12°17'12''.25 \pm 0''.4$ (Price, 2004) | 2000 |

| Detector: | Various AAVSO observer instruments. Details available upon request. |
|---|---|
| Filter(s): | CCD: $U, B, V, R_J, R_C; I_C$, unfiltered CCD |

| Date(s) of the observation(s): |
|---|
| 2004.09.11 − 2004.11.05 |

| Comparison star(s): | Finder chart and comparison stars are available at http://charts.aavso.org/. Comparison stars were based on the Tycho-2 catalog (comparison $V < 10^m.5$) and field photometry by Henden (comparison $V > 10^m.6$). |
|---|---|

| **Availability of the data:** |
|---|
| Data available for download at http://www.aavso.org/data/download |

| **Type of variability:** | UGWZ |
|---|---|

| **Remarks:** |
|---|
| ASAS 002511+1217.12 is a newly discovered cataclysmic variable in Pisces. It was discovered by G. Pojmanski and the ASAS-3 survey on 11.203 UT Sept., 2004 (Price, 2004). The AAVSO has collected 31839 CCD observations of ASAS 002511+1217.2 over a 55 day period following its discovery on Sep 11.203 (Price, 2004). Figure 1 shows all the CCD data. The errors depend on the observer and are available upon request but typically can be estimated to be $\pm\ 0\overset{m}{.}02$ for CCD observations. It is interesting to note that an echo-outburst occurred, reaching a maximum at JD 2453282.52. |
| Before combining data for statistical analysis, each observer's data set was individually transformed to an uniform zero-point by subtracting a linear fit from each night's observations. This was done so that we could remove the overall trend of outburst, and to combine all observations into a single data set. The analysis of CCD observations by the Lafler-Kinman (1965) method has enabled us to show the presence of the $0\overset{d}{.}05701 \pm 0\overset{d}{.}00006$ period of superhumps (Fig. 2). The periodogram presents the peak, which corresponds to the mentioned period. As an example of the superhump profile, we plotted the superhumps observed on Sept., 20, 2004 at the Crimean Astrophysical Observatory on the phase diagram of Fig. 3. |
| Taking JD 2453264.4332 as initial epoch for superhumps and period mentioned above, we build an $O - C$ diagram for superhump maxima (Fig. 4), using 71 times of superhump maxima. Not all nights' results are plotted on the $O - C$ diagram because some photometric data have too large scatter for accurate extrema determinations. The $O - C$ diagram presented here contains points between the super- and echo-outburst. Precise $O - C$ analysis at later time intervals (after echo-outburst) is complicated by the destruction of the superhump profile and, to a lesser extent, by increasing photometric noise due to the decreasing magnitude of the object. The solid line is an approximation by a 6th-order polynomial fit. |
| We propose that the period is not constant, but variable, as can be seen from the $O - C$ diagram. On the first 150 epochs the period increased with $\dot{P}$ ($d$P/P) = $4.77 \cdot 10^{-6} \pm 0.33 \cdot 10^{-6}$. Behavior of the period on further time scales should be investigated by future researchers, when more data become available. |

| **Acknowledgements:** |
|---|
| We acknowledge the help of AAVSO observers in continued monitoring of this object. This research has made use of the SIMBAD database, operated at CDS, Strasbourg, France. One of us ($PN^7$) would like to thank Curry Foundation and AAVSO for providing the $ST8XE$ camera and filters. ($AG^1$) is very grateful to Kira Makogon and Mykolaj Khotyaintsev for useful help and discussion on the question of the preparing this manuscript in LaTeX-format. |

References:

Price, A., 2004, *IAUC*, **8410**
Lafler, J., and Kinman, T.D., 1965, *Ap. J. Suppl.*, **11**, 216

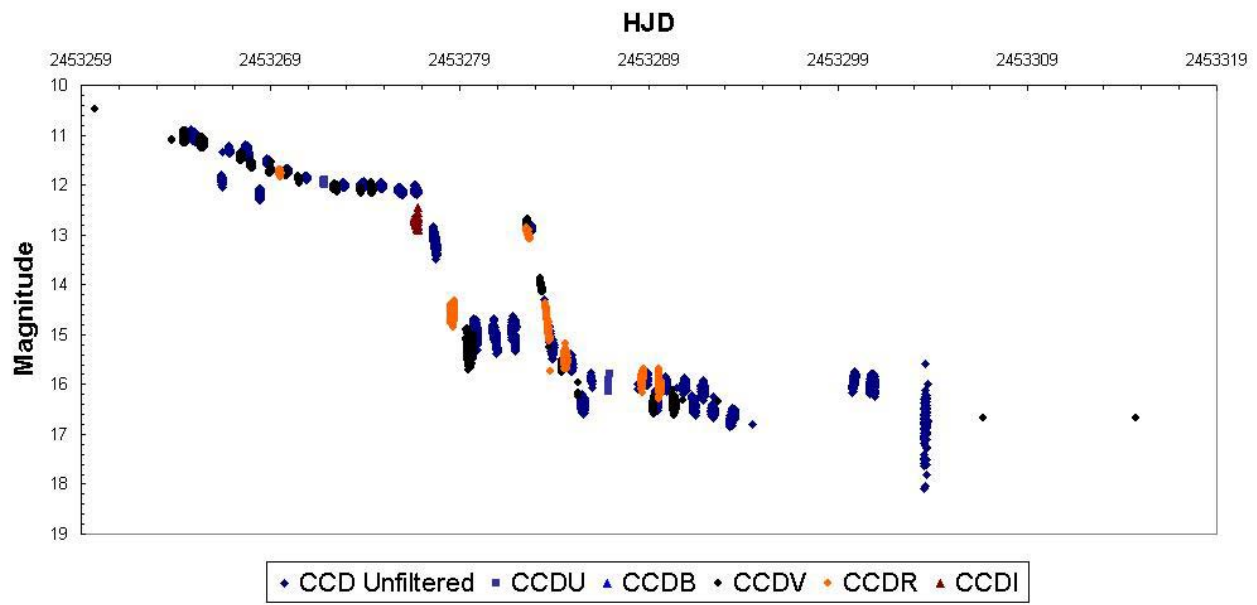

Figure 1. CCD Data

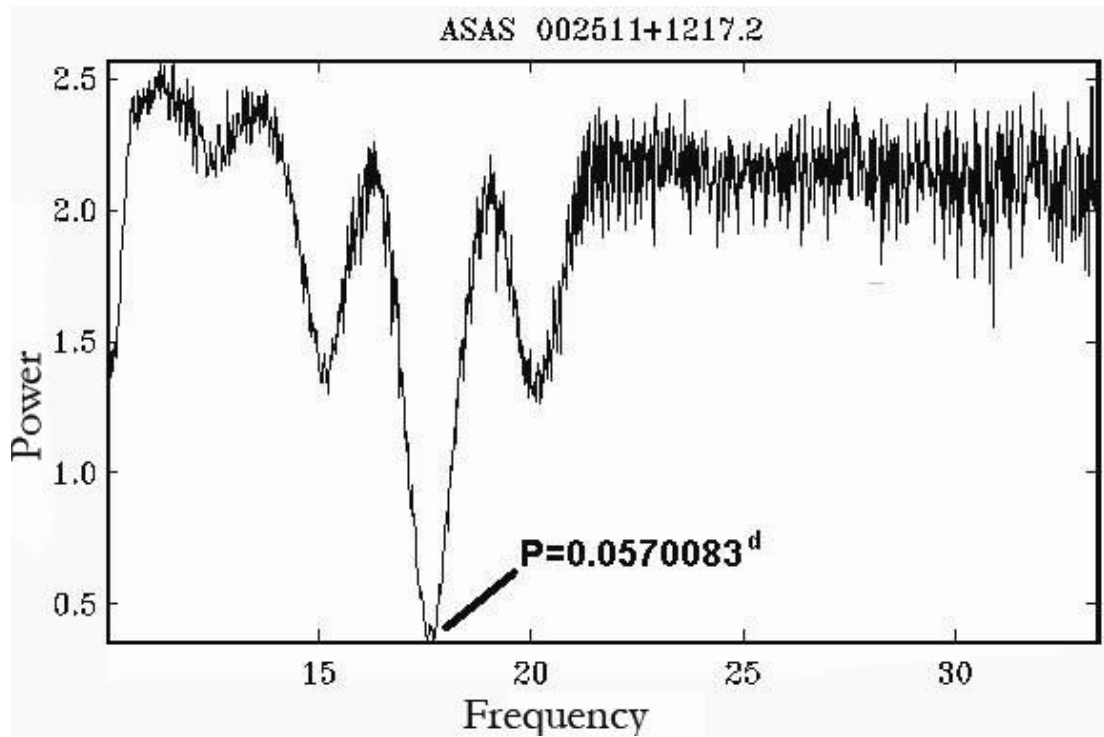

Figure 2. Periodogram

4                                                                                                          IBVS 5611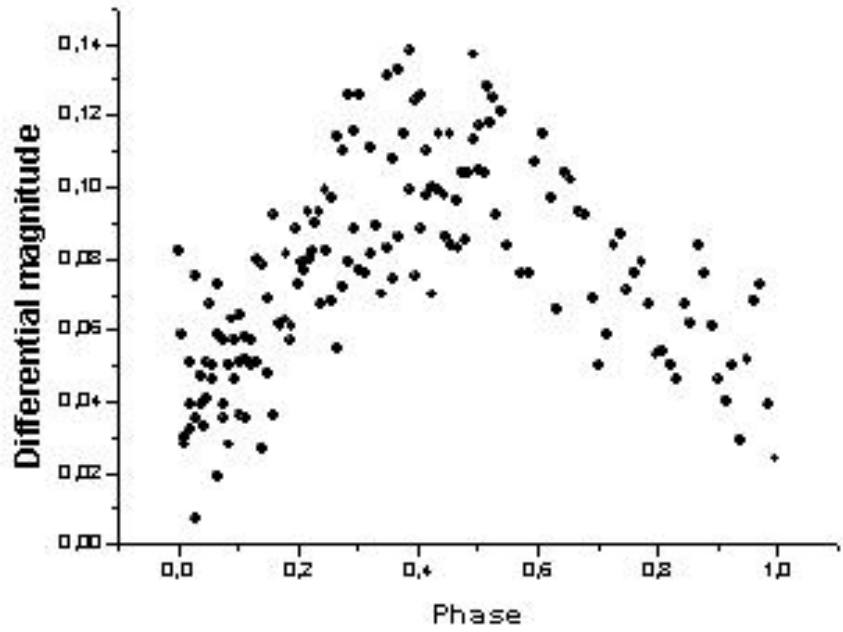

**Figure 3.** Phase diagram for superhumps

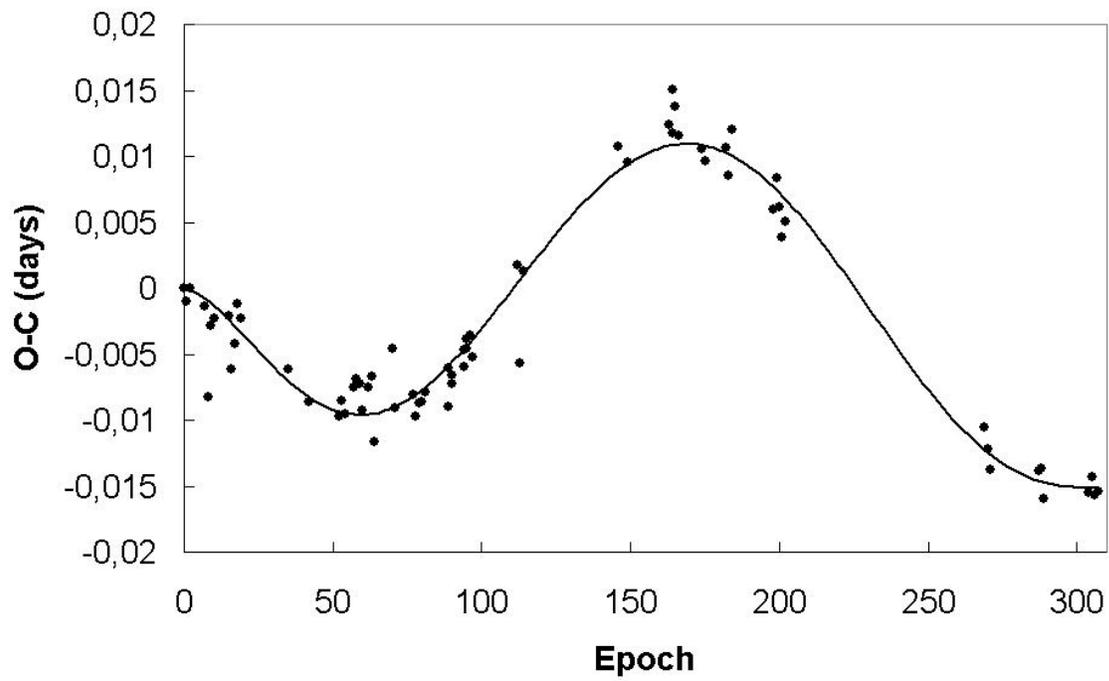

**Figure 4.** $O - C$